\documentstyle[12pt]{article}

\begin{document}
\thispagestyle{empty}
\baselineskip=20pt
\setcounter{page}{0}
\ \ \ 
\vspace{2.0cm}
\begin{center}
{\large \bf  Induced Gauge Structure of Quantum Mechanics on ${\bf S^D}$ \\}
\vspace{1.5cm}
Minoru HIRAYAMA, Hui-Min ZHANG and Takeshi HAMADA$^\ast$ \\
\vspace{1.5cm}
{\it Department of Physics, Toyama University, Toyama 930 \\
$\ast$Department of Physics, Kanazawa University, Kanazawa  920-11 }
\end{center}

\vspace{0.5cm}
\begin{center}
{\small (Received December 16,1996)}
\end{center}

\vspace{3cm}

{\small The Ohnuki-Kitakado (O-K) scheme of quantum mechanics on $S^D$ embedded in ${\bf \it R}^{D+1}$ is investigated. Generators satisfying the O-K algebra are written down explicitly in term of the induced gauge potential. A direct method is developed to obtain the generators in covariant form. It is seen  that there exists an induced gauge configuration which is trivial on $S^D$ but might cause a nontrivial physical effect in ${\bf \it R}^{D+1}$. The relation of the O-K scheme to extended objects such as the 't Hooft-Polyakov monopole is discussed}.

\newpage
\begin{center}
{\large \bf \S 1.  Introduction}\par
\end{center}
\vspace{0.5cm}

Recently several authors have discussed the manner in which  quantum mechanics should be formulated on manifolds.$^{1)-3)}$ It turns out that, on a manifold, there exist some inequivalent quantization schemes,  which can be identified with superselection sectors of the system. It is quite interesting that they could also discuss the origin of spin and gauge structure. Landsman and Linden,$^{3)}$  developing the canonical group quantization of Isham,$^{2)}$ discussed quantum mechanics on coset spaces. Regarding a circle $S^1$ and a sphere $S^2$ as ${\bf \it R}/{\bf \it Z}$ and $SO(3)/SO(2)$, respectively, they found that the Aharonov-Bohm $\theta$-angle and the Dirac magnetic monopole appear naturally in their quantum mechanics. Subsequently and independently, Ohnuki and Kitakado(O-K) investigated quantum mechanics on $S^D$ embedded in ${ \bf \it R}^{D+1}$.$^{4),5)}$   They found that the same  gauge structures as those of  Ref.3)  emerge in the representation of their fundamen!
tal algebra of observables on $S^D$. They also observed that, in an appropriate limit, their quantum mechanics on $S^D$ reduces to the  usual quantum mechanics on ${\bf \it R}^D$ with the spin automatically built into the theory. On the other hand, McMullun and Tsutsui developed a generalized version of Dirac's quantization of a constrained system$.^{6),7)}$  Regarding $S^4$ as ${\rm Spin}(5)\slash {\rm Spin}(4)$, they found that their $H$-connection reproduces a background BPST 
instanton and that the relativistic spin structure naturally arises in their quantization scheme. It should be noted that the induced gauge fields found in Ref.5) is in fact the $H$-connection.$^{8)}$

In this paper, we follow the line of thought of O-K.
Our discussion is made on the basis of their fundamental algebraic relations of observables. We develop a method which does not rely on Wigner's method of the little group. Discussion becomes simpler, in our opinion, and we directly obtain  results in  covariant form. We obtain a formula to express the O-K generator explicitly  in term of the   gauge potential and field strength. A gauge fixing condition leading us to the Wu-Yang Ansatz$^{9),10)}$ for magnetic monopoles and the Belavin-Polyakov-Schwartz-Tyupkin (BPST) ansatz$^{11)}$ for instantons is given. Our discussion reveals that there exist three classes of  solutions of the O-K algebra for any $D\geq 2$. Two of these three classes  of solutions yield vanishing field strength on $S^D$ and should be considered as trivial configurations on $S^D$. We discuss, however, that one of them might cause a nontrivial physical effect in ${\bf \it R}^{D+1}$ because of its sigularity at the origin of ${\bf \it R}^{D+1}$.  We also cons!
ider  how O-K's  quantum mechanics  on $\ S^D \ $ can include the physically allowed gauge configurations such as the  't Hooft-Polyakov monopole,$^{12),13)}$ Prasad-Sommerfield monopole,$^{14)}$ etc. We find that, if we wish  to include the extended objects mentioned above, the radius of  $S^D$ should be taken much larger than the size of the object concerned.

This paper is organized as follows. In $\S2$, we consider the general structure of the operator introduced by O-K. In $\S3$, we impose a gauge condition and find the solutions of the O-K algebra. In $\$ \S4$  and $5$,  we discuss the cases  $D=2$ and $D= 3$, respectively. The final section, $\S6$, is devoted to summary.\par

\vspace{3.5cm}
\begin{center}
{\large \bf $\S 2$. Structure of ${\it G}_{\alpha\beta}$}\par
\end{center}
  \vspace{0.5cm}
  
  The $D$-dimensional sphere $S^D$ embedded in ${\bf \it R}^{D+1}$ is defined by $x_\alpha\in {\bf \it R}, \alpha=1,2,\ \cdots, D+1,$ satisfying 
  $$ \sum_{\alpha=1}^{D+1}(x_\alpha)^2=r^2,  \eqno(2\cdot1) $$
  $$ [x_\alpha, x_\beta]=0, \eqno(2\cdot2)$$
where $r$ is a positive constant. O-K$^{5)}$ postulated that the fundamental algebra of quantum mechanics on $S^D\subset {\bf \it R}^{D+1}$ is given by 
$$ [x_\lambda , G_{\alpha\beta}]=i(x_\alpha\delta_{\lambda\beta}-x_\beta\delta_{\lambda\alpha}), \eqno(2\cdot3) $$
$$ [G_{\alpha\beta},G_{\lambda\mu}]=i(\delta_{\alpha\lambda}G_{\beta\mu}-\delta_{\alpha\mu}G_{\beta\lambda}+\delta_{\beta\mu}G_{\alpha\lambda}-\delta_{\beta\lambda}G_{\alpha\mu}), \eqno(2\cdot4)$$
where $G_{\alpha\beta}, \alpha,\beta=1,2,\\cdots,D+1,$ are self-adjoint operators. They sought the irreducible unitary representations of $G_{\alpha\beta}$ with the aid of Wigner's technique to obtain the representation of the Poincar${\acute e}$ group. Expressing $G_{\alpha\beta}$ as
 $$ G_{\alpha\beta}=L_{\alpha\beta}+f_{\alpha\beta}(x)=-G_{\alpha\beta},\eqno(2\cdot5)$$
 $$L_{\alpha\beta}=\frac1i(x_\alpha\partial_\beta-x_\beta\partial_\alpha),\eqno(2\cdot6)$$
O-K introduced the gauge potential $A_\alpha(x)$ by
$$ r^2A_\alpha(x)=\sum_{\beta=1}^{D+1}f_{\alpha\beta}x_\beta. \eqno(2\cdot7) $$
From $(2\cdot7)$ and $f_{\alpha\beta}(x)=-f_{\beta\alpha}(x)$, we have 
$$ \sum_{\alpha=1}^{D+1}x_\alpha A_\alpha (x)=0. \eqno(2\cdot8)$$
The gauge transformation on $S^D$ is caused by a unitary matrix $U(x)$ which is a representation of $SO(D+1)$ and satisfies 
$${\cal D}U(x)=0.  \eqno(2\cdot9)$$
Here $ {\cal D} $ is the dilation operator defined by
$${\cal D}=\sum_{\alpha=1}^{D+1}x_\alpha\partial_\alpha. \eqno(2\cdot10) $$
The condition $(2\cdot8)$ is preserved under such a gauge transformation:
${ \sum_{\alpha=1}^{D+1}}x_\alpha A^U_\alpha(x)=0, $ $ A^U_\alpha(x)=U(x)A_\alpha(x)U^\dagger(x)+iU(x)\partial_\alpha U^\dagger(x).$

To see what kind of gauge potential is allowed in the above scheme, we first obtain a formula expressing $f_{\alpha\beta}(x)$  by $A_\alpha(x)$. Although Eq. $(2\cdot7)$ cannot be solved algebraically  w.r.t. $f_{\alpha\beta}(x)$, it is possible to obtain the desired formula in the following way. Substituting $(2\cdot5)$ into  $(2\cdot4)$, we are led to 
$$ P_{\alpha\beta,\lambda\mu}[f]=i(x_\alpha\partial_\beta f_{\lambda\mu}-x_\beta \partial_\alpha f_{\lambda\mu})-i(x_\lambda\partial_\mu f_{\alpha\beta}-x_\mu \partial_\lambda f_{\alpha\beta} ), \eqno(2\cdot11)$$
where $P_{\alpha\beta,\lambda\mu}[f]$ is defined by 
$$ P_{\alpha\beta,\lambda\mu}[f]=[f_{\alpha\beta}, f_{\lambda\mu}]-i(\delta_{\alpha\lambda}f_{\beta\mu }-\delta_{\alpha\mu}f_{\beta\lambda}+\delta_{\beta\mu}f_{\alpha\lambda}-\delta_{\beta\lambda}f_{\alpha\mu}). \eqno(2\cdot12)$$
Multiplying $(2\cdot12)$ by $x_\mu x_\beta$, summing over $\mu$ and $\beta$, making use of $(2\cdot7)$ and $(2\cdot8)$, and noting the relations ${ \sum_{\mu=1}^{D+1}}({\cal D}f_{\lambda\mu})x_\mu = {\cal D}(r^2A_\lambda)-r^2A_\lambda$ and ${ \sum_{\mu=1}^{D+1}}(\partial_\alpha f_{\lambda\mu})x_\mu=\partial_\alpha(r^2A_\lambda)-f_{\lambda\alpha}$, we obtain
$$r^4[A_\alpha, A_\lambda]+ir^2f_{\alpha\lambda}=i\{x_\alpha {\cal D}(r^2A_\lambda)-x_\lambda {\cal D}(r^2A_\alpha)-r^2\partial_\alpha(r^2A_\lambda)+r^2\partial_\lambda(r^2A_\alpha)\}. \eqno(2\cdot13) $$
 We see that the condition $(2\cdot8)$ yields 
$$A_\alpha+{\cal D}A_\alpha=-\sum_{\beta=1}^{D+1}F_{\alpha\beta}x_\beta, \eqno(2\cdot14) $$
where $F_{\alpha\beta}(x)$ is the field strength defined by 
$$ F_{\alpha\beta}(x)=i[D_\alpha, D_\beta]=\partial_\alpha A_\beta(x)-\partial_\beta A_\alpha(x)-i[A_\alpha(x), A_\beta(x)],  \eqno(2\cdot15)$$
with $D_\alpha$ being the covariant derivative
$$ D_\alpha=\partial_\alpha-iA_\alpha(x). \eqno(2\cdot16)$$
We now obtain from $(2\cdot13),\ (2\cdot14)$ and $(2\cdot15)$ that 
$$ f_{\alpha\beta}(x)=-\{x_\alpha A_\beta(x)-x_\beta A_{\alpha}(x)\}+H_{\alpha\beta}(x), \eqno(2\cdot17) $$
$$ H_{\alpha\beta}(x)=-\sum_{\gamma=1}^{D+1}x_\gamma \{F_{\alpha\beta}(x)x_\gamma+F_{\beta\gamma}(x)x_\alpha +F_{\gamma\alpha}(x)x_\beta\}. \eqno(2\cdot18) $$
Equations $(2\cdot5)$ and $(2\cdot17)$ lead us to  
$$ G_{\alpha\beta}=M_{\alpha\beta}+H_{\alpha\beta}(x), \eqno(2\cdot19) $$
where $M_{\alpha\beta}$ is defined by
$$ M_{\alpha\beta}=\frac1i(x_\alpha D_\beta-x_\beta D_\alpha). \eqno(2\cdot20) $$
Since $D_\alpha$ and $F_{\alpha\beta}(x)$ are gauge covariant, the gauge covariance of $G_{\alpha\beta}$ is manifest in $(2\cdot19)$. We thus understand that the fundamental algebraic relations $(2\cdot 1)\sim (2\cdot 4)$ are gauge invariant.

We note that $G_{\alpha\beta}$ can be expressed solely by $M_{\alpha\beta}$ as follows. By definitions $(2\cdot12)$ and $(2\cdot 20)$, we have 
$$ P_{\alpha\beta,\lambda\mu}[M]=i(x_\alpha x_\lambda F_{\beta\mu}-x_\alpha x_\mu F_{\beta\lambda}+x_\beta x_\mu F_{\alpha\lambda}-x_\beta x_\lambda F_{\alpha\mu}). \eqno(2\cdot21) $$
Putting $\beta=\mu$ in $(2\cdot21)$ and summing over $\mu$, we obtain 
$$ \sum_{\mu=1}^{D+1}[M_{\alpha\mu}, M_{\lambda\mu}]-i(D-1)M_{\alpha\lambda}=-iH_{\alpha\lambda}(x). \eqno(2\cdot22)$$
Equations $(2\cdot19)$ and $(2\cdot22)$ yield 
$$ G_{\alpha\beta}=DM_{\alpha\beta} +i\sum_{\mu=1}^{D+1}[M_{\alpha\mu}, M_{\beta\mu}]. \eqno(2\cdot23)$$\par

\vspace{2.0cm}
\begin{center}
{\large \bf \S 3. Solutions of the fundamental algebra}\par
\end{center}
\vspace{0.5cm}
 In this section, we seek solutions of the algebraic relations $(2\cdot 1)\sim (2\cdot 4)$ . We denote the representation of the generators of the gauge group $SO(D+1)$ by $S_{\alpha\beta}=-S_{\beta\alpha}, \ \alpha, \ \beta=1,\ 2,\ \\cdots,\ D+1.$ They satisfy 
$$P_{\alpha\beta,\lambda\mu}[S]=0 \eqno(3\cdot1) $$
 and can be normalized as 
$$ tr(S_{\alpha\beta}S_{\lambda\mu})=\sigma (\delta_{\alpha\lambda}\delta_{\beta\mu}-\delta_{\alpha\mu}\delta_{\beta\lambda}), \eqno(3\cdot2) $$
 where $\sigma$ is a positive constant independent of $\alpha,\  \beta,\ \lambda$ and $\mu$. The gauge potential $A_{\alpha}(x)$ is written,  without loss of generality, as 
$$  A_{\alpha}(x)=\sum_{\beta,\gamma, \delta=1}^{D+1}E_{\alpha\beta\gamma\delta}(x)x_{\beta}S_{\gamma\delta}, \eqno(3\cdot3) $$
where $E_{\alpha\beta\gamma\delta}(x)$ is a function satisfying $E_{\alpha\beta\gamma\delta}(x)=- E_{\alpha\beta\delta\gamma}(x)$. To fix the transformation properties of $E _{\alpha\beta\gamma\delta}(x)$ under the coordinate transformation $x_\alpha\to x'_\alpha={ \sum_{\beta=1}^{D+1}} \Lambda_{\alpha\beta}x_{\beta},\ \Lambda=(\Lambda_{\alpha\beta})\in SO(D+1)$, we must fix that of $S_{\alpha\beta}$. Here we require  so that $S_{\alpha\beta}$ transforms as 
$$  S'_{\alpha\beta}=\sum_{\gamma,\delta=1}^ {D+1}\Lambda_{\alpha\gamma}\Lambda_{\beta\delta}S_{\gamma\delta}. \ \ \eqno(3\cdot4) $$
Then the vector property $A'_{\alpha}(x')\equiv{ \sum_{\beta,\gamma,\delta =1}^{D+1}} E'_{\alpha\beta\gamma\delta}(x')x'_{\beta}S'_{\gamma\delta}={ \sum_{\beta=1}^{D+1}}\Lambda_{\alpha\beta}A_{\beta}(x)$ indicates that $E_{\alpha\beta\gamma\delta}(x)$ is a fourth-rank tensor: $ E'_{\alpha\beta\gamma\delta}(x')={\sum_{\kappa,\rho,\lambda,\sigma=1}^{D+1}}\Lambda_{\alpha\kappa}\Lambda_{\beta\rho}\Lambda_{\gamma\lambda}\Lambda_{\delta\sigma}E_{\kappa\rho\lambda\sigma}(x).$  The structure of the tensor $E_{\alpha\beta\gamma\delta}(x)$ should be fixed by the condition $(2\cdot 8)$ and a gauge   fixing condition. We impose the following gauge fixing condition
$$ \sum_{\beta,\gamma=1}^{D+1} [\{{\rm tr}(S_{\alpha\beta}S_{\beta\gamma})\}A_\gamma(x)-S_{\alpha\beta}tr\{S_{\beta\gamma}A_\gamma(x)\}]=0. \eqno(3\cdot5) $$
We can argue that, for any $A_\alpha(x)$ satisfying $(2\cdot8)$, there exists a gauge transformation $U(x)$ satisfying $(2\cdot9)$ such that $A_\alpha^U (x)$ obeys $(2\cdot8)$ and $(3\cdot5)$. From $(3\cdot2)$, $ (3\cdot3)$ and $(3\cdot5)$, we obtain 
$$
E_{\alpha\beta\gamma\delta}(x)=\delta_{\alpha\gamma}e_{\beta\delta}(x)-\delta_{\alpha\delta}e_{\beta\gamma}(x),$$
$$e_{\beta\delta}(x)={\displaystyle \frac1{D}\sum_{\alpha=1}^{D+1}} E_{\alpha\beta\alpha\delta}(x). \eqno(3\cdot6) $$
The condition $ (2\cdot8) $  then yields $e_{\beta\delta}(x)=J_{\beta}(x)x_{\delta}$, where $J_\beta(x)$ is a vector field. Putting the scalar ${ \sum_{\alpha=1}^{D+1}} x_\alpha J_\alpha(x)$  as ${\displaystyle \frac12}V(r),$ we are led to 
$$ \sum_{\beta=1}^{D+1}E_{\alpha\beta\gamma\delta}(x)x_\beta=-\frac12V(r)(x_\gamma\delta_{\alpha\delta}-x_\delta\delta_{\alpha\gamma}), \eqno(3\cdot7)$$
which is, for $D=3$,  equivalent to the ansatz adopted by BPST in their pioneering paper on the instanton.$^{11)}$  We stress that, in the prensent context associated with the condition $(2\cdot 8)$, the BPST Ansatz corresponds to the gauge condition $(3\cdot 5)$. From $(3\cdot 3)$ and $(3\cdot 7)$, we obtain 
$$
 A_\alpha(x)={\displaystyle  \frac12\sum_{\beta,\gamma=1}^{D+1}} A_\alpha^{\beta\gamma}(x)S_{\beta\gamma},$$
$$A_\alpha^{\beta\gamma}(x)=-V(r)(x_\beta\delta_{\alpha\gamma}-x_\gamma\delta_{\alpha\beta})  \eqno(3\cdot8) $$
and hence
$$ A_\alpha(x)= V(r)\sum_{\beta=1}^{D+1}S_{\alpha\beta}x_\beta. \eqno(3\cdot9) $$
It is evident that, for $D=2$, the expression $(3\cdot9)$ for $A_\alpha(x)$ coincides with that adopted by Wu and Yang.$^{9)}$  The field strength is now given by  
$$ F_{\alpha\beta}(x)=h(r) S_{\alpha\beta}+j(r)J_{\alpha\beta}(x), \eqno(3\cdot10) $$
where $h(r),\ j(r)\ $ and $J_{\alpha\beta}(x)$ are defined  by
$$ h(r)=r^2V(r)^2-2V(r),\eqno(3\cdot11) $$
$$ j(r)=r^2V(r)^2+rV'(r),\eqno(3\cdot12)$$
$$J_{\alpha\beta}(x)=\sum_{\gamma=1}^{D+1}({\hat x}_\alpha S_{\beta\gamma}-{\hat x}_\beta S_{\alpha\gamma}){\hat x}_\gamma, \ \ \ {\hat x}_\gamma=\frac {x_\gamma}r. \eqno(3\cdot13)$$
It is straightforward to obtain
$$ G_{\alpha\beta}=L_{\alpha\beta}-r^2h(r)S_{\alpha\beta}-r^2\{V(r)+h(r)\}J_{\alpha\beta},\eqno(3\cdot14)$$
where we have made use of the identity
$$ \sum_{\gamma=1}^{D+1}\{J_{\alpha\beta}(x)x_\gamma+J_{\beta\gamma}(x)x_\alpha+J_{\gamma\alpha}(x)x_\beta\}x_\gamma=0.\eqno(3\cdot15)$$
We see that $G_{\alpha\beta}$ is independent of the derivative of $V(r)$.

The commutation relations among $J_{\alpha\beta}(x),\ S_{\alpha\beta}$ and $L_{\alpha\beta}$ are calculated to be
$$ [J_{\alpha\beta}, J_{\gamma\delta}]=iK_{\alpha\beta,\gamma\delta} \eqno(3\cdot16) $$
$$ [S_{\alpha\beta}, J_{\gamma\delta}]=-iK_{\alpha\beta,\gamma\delta} -iN_{\alpha\beta,\gamma\delta}  \eqno(3\cdot17)$$
$$ [L_{\alpha\beta}, J_{\gamma\delta}]=iK_{\alpha\beta,\gamma\delta} +iN_{\gamma\delta,\alpha\beta} \eqno(3\cdot18) $$
where $K_{\alpha\beta,\gamma\delta}$ and $N_{\alpha\beta,\gamma\delta}$ are defind by 
$$ K_{\alpha\beta,\gamma\delta}={\hat x}_\alpha{\hat x}_\gamma S_{\beta\delta}-{\hat x}_\beta{\hat x}_\gamma S_{\alpha\delta}+{\hat x}_\beta{\hat x}_\delta S_{\alpha\gamma}-{\hat x}_\alpha{\hat x}_\delta S_{\beta\gamma},\eqno(3\cdot19)$$
$$N_{\alpha\beta,\gamma\delta}=({\hat x}_\gamma\delta_{\beta\delta}-{\hat x}_\delta\delta_{\beta\gamma})\sum_{\kappa=1}^{D+1} S_{\alpha\kappa}{\hat x}_\kappa-({\hat x}_\gamma\delta_{\alpha\delta}-{\hat x}_\delta\delta_{\alpha\gamma})\sum_{\kappa=1}^{D+1}S_{\beta\kappa}{\hat x}_\kappa.\eqno(3\cdot20) $$
We now obtain 
$$\begin{array}{l}
 \ \ P_{\alpha\beta, \gamma\delta}[G]\ \ \ \ \ \ \ \ \ \\
\ \ \ \ \ \ \ \ \ \\
 \ \ \ \ \ \ \ \ =\{(r^2h)^2+r^2h\}[S_{\alpha\beta}, S_{\gamma\delta}]-ir^2(h+V)(r^2h-r^2V+2)K_{\alpha\beta.\gamma\delta}\\
\ \ \ \\
 \ \ \ \ \ \ \ \ \ \ \ \ \ \ \ \   -ir^4(h+V)h(N_{\alpha\beta,\gamma\delta}-N_{\gamma\delta,\alpha\beta})\\
\ \ \ \\
\ \ \ \ \ \ \ \  = r^2V(r^2V-1)(r^2V-2)\{(r^2V-1)[S_{\alpha\beta},S_{\gamma\delta}]\end{array}   $$
 $$-i(r^2V-1)K_{\alpha\beta,\gamma\delta}
-ir^2V(N_{\alpha\beta,\gamma\delta}-N_{\gamma\delta,\alpha\beta})\}.\eqno(3\cdot21)$$
The requirement $(2\cdot4)$, i.e., $$
P_{\alpha\beta, \gamma\delta}[G]=0 \eqno(3\cdot22)$$
 yeilds the result
$$ r^2V\{r^2V-1\}\{r^2V-2\}=0.\eqno(3\cdot23) $$
 We thus  obtain three solutions 
$$ ({\rm  a})\ \ \ \ \ \ r^2V(r)=0, \eqno(3\cdot24\rm a) $$
$$ ({\rm  b})\ \ \ \ \ \ r^2V(r)=1, \eqno(3\cdot24\rm b) $$  
$$ ( {\rm c})\ \ \ \ \ \ r^2V(r)=2, \eqno(3\cdot24\rm c) $$
and  $G_{\alpha\beta}$ is given by 
$$ ( {\rm a})\ \ \ \ \ \ \ \ \ \ \ G_{\alpha\beta}=L_{\alpha\beta},\ \ \ \ \ \   \eqno(3\cdot25\rm a) $$
$$ ({\rm  b})\ \ \ \ \ \ \ G_{\alpha\beta}=L_{\alpha\beta}+S_{\alpha\beta},\ \eqno(3\cdot25\rm b)$$
$$ ( {\rm c})\ \ \ \ \ \ \ G_{\alpha\beta}=L_{\alpha\beta}-2J_{\alpha\beta}(x), \eqno(3\cdot25\rm c) $$
in the  respective cases. It can be seen that both  cases $(\rm a)$ and $(\rm c)$ yield vanishing field strengths on $S^D$. If we denote  case $(\rm b)$ with the singlet representation for $S_{\alpha\beta}$ by $(\rm b_0)$, the field strength for the case $(\rm b_0)$ is also vanishing on $S^D$. Regarding $(\rm a)=(\rm b_0)$ hereafter,  cases $(\rm b_0)$ and $(\rm c)$ are trivial as the induced gauge potential on $S^D$. It seems that in the O-K approach besed on Wigner's little group, the  case $(\rm c)$ is absorbed into  case $(\rm b_0)$ because they are gauge equivalent to each other. It should be noted, however, that the gauge configuration of $(\rm c)$ exhibits quite a different property from that of $(\rm b_0)$ at the origin of ${\bf \it R}^{D+1}$. We shall discuss later the manner in which  they differ.

\vspace{2cm}
\begin{center}
{\large \bf $\S 4.$ Magnetic monopole solution}
\end{center}
\vspace{0.5cm}
\par
 Here we investigate the case  $D=2$.  In contrast to the case  $D\geq 3$, at least two of $\alpha, \beta,\lambda$ and $ \mu$ in $P_{\alpha\beta,\lambda\mu}[G]$ coincide for $D=2$. Because of the anti-symmetry $G_{\alpha\beta}=-G_{\beta\alpha},$ it is sufficient to consider three cases in 
$P_{\alpha\beta,\lambda\mu}[G]=0:\ ($i$)\ \alpha=\mu=1,\ \beta=2,\ \lambda=3,\ ($ii$)\  \alpha=\mu=2, \ \beta=3,\ \lambda=1,\ ($iii$) \ \alpha=\mu=3,\ \beta=1,\ \lambda=2$. Equation  $(3\cdot21)$ simplifies to
$$ 
P_{\alpha\beta,\lambda\alpha}[G]=[H(x),{\displaystyle \sum_{\kappa=1}^3} x_\kappa M_{\alpha\kappa}] $$

$$(\alpha\beta\lambda)=(123),\ (231),\ (312),  \eqno(4\cdot1) $$
where $H(x)$ is given by 
$$ H(x)=x_1F_{23}(x)+x_2F_{31}(x)+x_3F_{12}(x). \eqno(4\cdot2)$$
For the sake of comparison  with earlier works, it is  convenient to use
$$ T_\gamma=\frac 12 \sum_{\alpha,\beta=1}^3\epsilon_{\gamma\alpha\beta}S_{\alpha\beta}, \eqno(4\cdot3\rm a) $$
$$ [T_\alpha,T_\beta]=i\sum_{\gamma=1}^3\epsilon_{\alpha\beta\gamma}T_\gamma             \eqno(4\cdot3\rm b) $$
instead of $S_{\alpha\beta}$. Equation $(3\cdot8)$ then becomes 
$$ A_\alpha(x)=\sum_{\beta,\gamma=1}^3 \epsilon_{\alpha\beta\gamma}x_\beta T_\gamma V(r), \eqno(4\cdot4)$$
which is nothing but the Wu-Yang ansatz$^{9)}$ for a  three-dimensional Yang-Mills field.
The function $H(x)$ in $(4\cdot2)$ is calculated to be 
$$ H(x)=\{r^2V(r)^2-2V(r)\}(\sum_{\gamma=1}^3 x_\gamma T_\gamma), \eqno(4\cdot5) $$
and we find that the r.h.s. of $(4\cdot1)$ is given by
$$ [H, \sum_{\gamma=1}^3x_\gamma M_{\alpha\gamma}]=-ir^2V(r)\{r^2V(r)-1\}\{r^2V(r)-2\}\{{\hat x}_\alpha(\sum_{\gamma=1}^3{\hat x}_\gamma T_\gamma)-T_\alpha\}.\ \ \eqno(4\cdot6) $$
We find that the $V(r)^4$-term in $(3\cdot21)$ cancels out in the $D=2$ case. In the following, we discuss the three solutions of $(3\cdot22)$ given in $(3\cdot24)$ and $(3\cdot25)$.

Solution  $(\rm a)$ is trivial and equivalent to $(\rm b_0)$ defined at the end of $\S3.$

Solution $(\rm b)$ corresponds to the one obtained by O-K.$^{5)}$
 The gauge potential in this case is the Wu-Yang solution of the pure Yang-Mills field theory. This configuration is known to be gauge equivalent to the following:$^{10)}$  
$$ 
A_\alpha(x)={\displaystyle \sum_{\gamma=1}^3}A_\alpha^\gamma(x)T_\gamma, $$
$$ A_\alpha^1=A_\alpha^2=0, \ \ A_\alpha^3=-{\displaystyle \frac1{2r}\tan(\frac12\theta){\bf \it e}_\phi},
\eqno(4\cdot7)$$
where $(\theta,\phi)$ is the polar coordinate on $S^2$, and ${\bf \it  e}_\phi$ is the unit vector in the  direction of $\phi$. As has been discussed by many people, this configuration  describes a gauge potential caused by a point-like magnetic monopole.$^{10)}$

We next consider the solution  $(\rm c)$,  which was not considered  by O-K.$^{5)}$  As we discussed in  the last paragraph of $\S 3,\ (\rm c)$ is gauge equivalent to $(\rm b_0)$. Since the field strength vanishes for $r > 0$ in this case, the gauge potential can be expressed as  a pure gauge in a simply connected domain which does not contain the origin: $A_\alpha=iU\partial _\alpha U^\dagger$. The operator $G_{\alpha\beta}$ is  given by $G_{\alpha\beta}=UL_{\alpha\beta}U^\dagger$ and we can check $(2\cdot 4)$ by $P_{\alpha\beta, \lambda\mu}[G]=UP_{\alpha\beta,\lambda\mu}[L]U^\dagger=0$. Although any $G_{\alpha\beta}$ of the above form satisfies $(2\cdot 4)$, we here obtain a highly specified form of $G_{\alpha\beta}$, $(3\cdot 25\rm c)$. This specification should be attributed to the gauge condition $(3\cdot5)$. We note that we can replace this conditopn  by 
$$ {\rm tr} \{(\sum_{\gamma=1}^3 x_\gamma T_{\gamma})A_\alpha(x)\}=0. \eqno(4\cdot8) $$
Although this configuration does not correspond to the magnetic monopole, it is nontrivial in ${\bf \it R}^3$ because of its singularity at the origin.
The unitary matrix $U$ for  case $(\rm c)$ is given by $ U=e^{i\pi S},\  S={\hat x}_1 T_1+ {\hat x}_2 T_2 +{\hat x}_3 T_3.$ The structure of the singularity at the origin can be envisaged by calculating  the quantity $Q$ defined by 
$$ Q=\int _{{\bf \it R}^3} \rho(x)d^3x=\int_{{\bf \it R}^3} \sum_{i=1}^3 \partial_i\xi_i(x)d^3x,   \eqno(4\cdot9)$$ 
$$ \rho(x)= i\sum_{i,j,k =1}^3 \epsilon_{ijk}tr(UA_i A_j A_k), \eqno(4\cdot10) $$
$$\xi_i(x)=\sum_{j,k=1}^3 \epsilon_{ijk}tr(UA_jA_k). \eqno(4\cdot11) $$
Although $\rho(x)$ vanishes for any $r>0$, the r.h.s. of $(4\cdot9)$ is equal to 
$ \ \ {\displaystyle 4i\int_{S^2}tr(e^{i\pi S}S)d\Omega}\ \ $ and nonvanishing in general, implying that $\rho(x) $ has a $\delta$-function singularity at the origin.

 The field equation  of the pure $SO(3)$ Yang-Mills theory under the Wu-Yang Ansatz $(4\cdot4)$ is given by$^{10)}$
$$r^2\frac {d^2}{dr^2}\{r^2V(r)\}=r^2V(r)\{r^2V(r)-1\}\{r^2V(r)-2\}. \eqno(4\cdot12) $$
It is interesting to note that the algebraic requirement $(2\cdot 4)$ reproduces all the  solutions of $(4\cdot 12)$ of the type $r^2V(r)={\rm const}.$

We have obtained in the above the gauge configuration of a point-like monopole. On the other hand, we know some examples of extended monopoles,   the 't Hooft-Polyakov$^{12),13)}$ monopole, the Prasad-Sommerfield$^{14)}$  monopole, etc., of the $SO(3)$ Yang-Mills-Higgs field theory. The gauge configurations corresponding to these examples still take the form of $(4\cdot4)$, but the function $V(r)$  in these cases does not satisfy the condition $(3\cdot 23)$. We find, however, that the function $r^2V(r)-1$  for the 't Hooft-Polyakov as well as the Prasad-Sommerfield monopoles decreases exponentially for large values of $r$:
$$ r^2V(r)-1\approx {\rm const}.r e^{-\beta r},\ \ (r\approx \infty) \eqno(4\cdot13)$$
 where $\beta^{-1} $ is the size parameter. Thus, instead of $P_{\alpha\beta,\lambda\alpha}[G]=0( \ \infty > r >0)$, we have 
 $$ 
P_{\alpha\beta,\lambda\alpha}[G]\approx {\rm const}.r e^{-\beta r}\{{\hat x}_\alpha({\displaystyle\sum_{\gamma=1}^3}{\hat x}_\gamma T_\gamma)-T_\alpha\},\ \  (r\approx  \infty)$$
$$ (\alpha\beta\lambda)=(123),\ (231),\ (312). \eqno(4\cdot14) $$
In other words, the condition
$$P_{\alpha\beta,\lambda\alpha}[G]\approx 0,\ \ \ (r \gg\beta^{-1} >0) \eqno(4\cdot15) $$
allows for  gauge configurations of the extended monopole of the above type. It should be noted that the Higgs field is concerned with the dynamics of a particle on $S^2$ but not with its kinematics. Since the fundamental algebra should be independent of the dynamics, only the Yang-Mills field appeared in the above discussion. Of course, the details of the gauge configuration cannot be determined only through $(4\cdot15)$.\par

\vspace{2cm}
\begin{center}
{\large \bf $\S 5$. BPST instanton solution}
\end{center}
\par
\vspace{0.5cm}
In this section, we consider the case $D=3,$ i.e., the O-K algebra for $S^3$ embedded in ${\bf \it R}^4$. Results for $V(r)$ and $G_{\alpha\beta}$ are given by $(3\cdot24)$ and $(3\cdot25)$.  Case $(\rm a)$ is trivial and identical with $(\rm b_0)$, as  mentioned in $\S3$. Solution $(\rm b)$ was   discussed by O-K.$^{5)}$  For this solution, we see that ${\rm tr}\{{\sum_{\mu,\nu=1}^4}(F_{\mu\nu})^2\}$ is proprotional to $r^{-4}$,  and the corresponding action integral is divergent. Its singularity structure at the origin was discussed by O-K in detail.

On the other hand, we have 
$$ F_{\alpha\beta}(x)=0,\ \ \ \ \ (r > 0) \eqno(5\cdot1) $$
for the solution $(\rm c)$. Comparing with BPST,$^{11)}$   however, this solution should be interpreted as the zero size limit of the BPST solution. To understand the above interpretation, we replace $V(r)=2r^{-2}$ by $V_\lambda(r)=2(r^2+\lambda^2)^{-1}$, where $\lambda$ is the size pamameter  which can be taken as small as desired. The field strength then becomes      
$$ F_{\alpha\beta}^\lambda(x)=-\frac{4\lambda^2}{(r^2+\lambda^2)^2}S_{\alpha\beta},  \eqno(5\cdot2) $$
 which is  the configuration considered by BPST.$^{11)}$ Another way of understanding the above interpretation is to calculate the $SU(2)$ instanton number, $q$ ,  corresponding to the configuration $A_\alpha^{\beta\gamma}(x)=-2(x_\beta\delta_{\alpha\gamma}-x_\gamma\delta_{\alpha\beta})r^{-2}.$  Faithfully following the method of BPST,$^{11)}$ we obtain $q=\pm 1$.
 Two values,  $+1$ and $-1$,  for  $q$ are allowed because there are two ways to reduce the $SO(4)$ gauge potential to the $SU(2)$ gauge potential.  We expect that this configuration might cause a nontrivial effect for physics in ${\bf \it R}^4$ and the instanton number $q$ plays a similar role to that of the thin magnetic flux in the Aharonov-Bohm effect.

We note here  some differences between previous works and those prensented in this paper. Fujii, Kitakado and Ohnuki$^{15)}$ considered quantum mechanics on $S^{2n}  , n=2,3,\\cdots,$ embedded in ${\bf \it R}^{2n+1}$. They stereographically projected their $2n+1$ dimensional gauge potential to that of $S^{2n}$ and obtained a generalized BPST configuration$^{16)}$  with a nonvanishing scale parameter. On the contrary, we considered the  O-K algebra for $S^3$ in ${\bf \it R}^4$ and obtained the BPST confuguration in $ {\bf \it R}^4$ with a vanishing size parameter. We here encounter a situation similar to that of the previous section: the O-K algebra excludes extended objects. If we adopt $(4\cdot15)$ instead of $(2\cdot4)$, we are allowed to include the BPST configuration with  finite size.    \par

\vspace{2cm}
\begin{center}
{\large \bf $\S 6$. Summary}
\end{center}
\par
\vspace{0.5cm}
We have investigated the representation of the Ohnuki-Kitakado algebra $(2\cdot 1)\sim (2\cdot 4)$ for quantum mechanics on $S^D$ embedded in ${\bf \it R}^{D+1}$ without using  Wigner's little group method. The function $f_{\alpha\beta}(x)$ in $(2\cdot 5)$ was represented by $(2\cdot17)$ in terms of $A_\alpha(x)$ and $F_{\alpha\beta}(x)$. The expressions $(2\cdot 19)$ and $(2\cdot 23)$ for $G_{\alpha\beta}$ manifestly exhibit its gauge covariance, implying the gauge invariance of the O-K algebra. We have observed that the gauge condition $(3\cdot 5)$ naturally leads us to the Wu-Yang ansatz for magnetic monopoles and the BPST ansatz for pseudoparticle solutions. 
Three classes of  solutions of the O-K algebra were obtained: $(\rm a),\ (\rm b)$ and $(\rm c)$. Class $(\rm a)$ which is identical to the $(\rm b_0)$, the singlet representation case of $(\rm b)$, is trivial both on $S^D$ and ${\bf \it R}^{D+1}$ . Class $(\rm b)$ is the one discussed by O-K and nontrivial on $S^D$. The field strength for the class $(\rm c)$ vanishes on $S^D$, implying that the configuration is trivial on $S^D$. Configurations belonging to this class, however, might produce some physical effects  in ${\bf \it R}^{D+1}$. It was noted for the case $D=2,\ 3$ that the $Q$ of $(4\cdot9)$ and the instanton number $q$ might play the role  similar to that of the thin magnetic flux in the Aharonov-Bohm effect.
We have also discussed how the gauge configurations of extended objects such as the 't Hooft-Polyakov monopole can be included in the O-K scheme of quantum mechanics in $S^D$.
\par      

\vspace{2.0cm}
\begin{center}
{\bf { Acknowledgements}}
 \end{center}
\par
\vspace{0.8cm}
 One of the authors (M.H.) is grateful to Professor  Z. Z. Xin for his hospitality at Liaoning University, China, where a part of this paper was written.

\newpage
{\small
\vspace{2.5cm}
\begin{center}
{ \bf References}
\end{center}
\par
\vspace{0.8cm}
1) G. W. Mackey, {\it Induced Representations of Groups And Quantum Mechanics} 

\ \ \ \ (Benjamin, New York, 1969).

2) C. J. Isham, in {\it Relativity, Groups And Topology II}, ed. B. S. DeWitt and R. Stora 

\ \ \ \ (North-Holland, Amsterdam, 1984).

3) N. P. Landsman amd N. Linden, Nucl. Phys. {\bf B365} (1991), 121.

4) Y. Ohnuki and S. Kitakado, Mod. Phys. Lett. {\bf A7} (1992), 2477.

5) Y. Ohnuki and S. Kitakado, J. Math. Phys. {\bf 34} (1993), 2827.

6) D. McMullan and I Tsutsui, Phys. Lett. {\bf B320} (1994), 287.

7) D. McMullan and I Tsutsui, Ann. of  Phys. {\bf 237} (1995), 269.

8) P. L${\acute e}$vay, D. McMullan and I. Tsutsui, J. Math. Phys. {\bf 37} (1996), 625.

9) T. T. Wu and C. N. Yang, in {\it Properties of Matter under Unusual Conditions},

\ \ \ \  ed. H. Mark and S.Fernbach (Interscience, New York, 1968).

10) A. Actor, Rev. Mod. Phys. {\bf 51} (1979), 461.

11) A. A. Belavin, A. M. Polyakov, A. S. Schwartz and Yu. S. Tyupkin, Phys. Lett.

\ \ \ \ {\bf 59B} (1975), 85.

12) G. 't Hooft, Nucl. Phys. {\bf B79} (1974), 276.

13) A. M. Polyakov, JETP Lett. {\bf 20} (1974), 194.

14) M. K. Prasad and C. M. Sommerfield, Phys. Rev. Lett. {\bf 35} (1975), 760.

15) K. Fujii, S. Kitakado and Y. Ohnuki, Mod. Phys. Lett. {\bf A10} (1995), 867.

16) K. Fujii, Lett. Math. Phys. {\bf 12} (1986), 363.

}

\par

\end{document}